\documentstyle [12pt,epsfig]{article} 
\makeatletter

\newcommand{\CAP}{{\rm C}'_{A}}
\newcommand{\CBP}{{\rm C}'_{B}}
\newcommand{\rC}{{\rm C}}

\@addtoreset{equation}{section}
\makeatother

\newcommand{\ba}{\begin{eqnarray}}
\newcommand{\ea}{\end{eqnarray}}

\newcommand{\MA}{$\mu_{\rm A}$ }
\newcommand{\MB}{$\mu_{\rm B}$ }
\newcommand{\MAP}{$\mu_{\rm A'}$ }

\begin{document}
\newcommand{\BS}{\bigskip}
\newcommand{\SECTION}[1]{\BS{\large\section{\bf #1}}}
\newcommand{\SUBSECTION}[1]{\BS{\large\subsection{\bf #1}}}
\newcommand{\SUBSUBSECTION}[1]{\BS{\large\subsubsection{\bf #1}}}

\begin{titlepage}
\begin{center}
\vspace*{2cm}
{\large \bf Muon decays in the Earth's atmosphere, differential aging and the paradox
  of the twins}  
\vspace*{1.5cm}
\end{center}
\begin{center}
{\bf J.H.Field}
\end{center}
\begin{center}
{ 
D\'{e}partement de Physique Nucl\'{e}aire et Corpusculaire
 Universit\'{e} de Gen\`{e}ve . 24, quai Ernest-Ansermet
 CH-1211 Gen\`{e}ve 4.
}
\newline
\newline
   E-mail: john.field@cern.ch
\end{center}
\vspace*{2cm}
\begin{abstract}
   Observation of the decay of muons produced in the Earth's atmosphere by
    cosmic ray interactions provides a graphic illustration of the
    counter-intuitive space-time predictions of special relativity theory.
    Muons at rest in the atmosphere, decaying simultaneously, are
    subject to a universal time-dilatation effect when viewed from a
    moving frame and so are also
    observed to decay simultaneously in all such frames. The analysis of
   this example reveals the underlying physics of the differential aging
   effect in Langevin's travelling-twin thought experiment.
 \par \underline{PACS 03.30.+p}

\vspace*{1cm}
\end{abstract}
\end{titlepage}
 
\SECTION{\bf{Introduction}}
    The present paper is one in a recent series, devoted to space-time physics, by the
   present author.
   In Ref.~\cite{JHFRSPB} the classical `Rockets-and-String'
    ~\cite {RS} and `Pole-and-Barn'~\cite{PB} paradoxes of special relativity were re-analysed taking
    into account the distinction between the real and apparent\footnote{i.e. as naively predicted by 
   the standard space-time Lorentz transformation} positions of uniformly moving objects. Different
    results were obtained from those of the usual text-book interpretations of these experiments and a new
    causality-violating paradox was pointed out. This paradox, as well as the related `backwards running
    clocks' one of Soni~\cite{Soni}, was resolved in Ref.~\cite{JHFLLT} where, in order to avoid these
     paradoxes as well as manifest
     breakdown of translational invariance, in some applications of the standard space-time Lorentz
    transformation, the use of a `local' Lorentz transformation. i.e. one where the transformed event in
    the moving frame lies at the coordinate origin in this frame, was advocated. When this is done the
    closely correlated `relativity of simultaneity' (RS) and `length contraction' (LC) effects of 
    conventional special relativity theory do not
    occur. The connection between these effects is explained in Ref.~\cite{JHFRSPB}.
      Ref.~\cite{JHFLLT} also contains a `mini review' of all experimental tests of special relativity
      where it is pointed out that, whereas time dilatation is well-confirmed experimentally,
     no experimental evidence exists for the RS and LC effects. In the later papers
     ~\cite{JHFUMC,JHFCRCS,JHFACOORD} it is explained how the spurious RS and LC effects result from a
     misuse of the time symbols in the standard space-time Lorentz transformation.
     Refs.~\cite{JHFCRCS,JHFACOORD} present
    the argument in a pedagogical manner, whereas Ref.~\cite{JHFUMC} contains a concise summary of it.
   \par In the following section the necessary formulae for the analysis of the muon decay thought
    experiment ---essentially the prediction of a universal time dilatation effect--- are derived from
    first principles. Here there is considerable overlap with work presented in
   Refs.~\cite{JHFUMC,JHFCRCS,JHFACOORD}
    The analysis of the thought experiment presented in Section 3 shows the
    absence of the spurious text-book RS effect: Muons which decay simultaneously
     in a common proper frame, are observed to do so in all inertial frames. Finally the results
     of the analysis presented in this paper are used to shed light on the physical basis
     of the differential aging effect in the travelling-twin thought experiment~\cite{Langevin}.

\SECTION{\bf{Operational meaning of the space-time Lorentz transformation: Rates and spatial
separations of moving clocks}}
 The Lorentz transformation (LT) relates observations ($x$,$y$,$z$,$t$) of the coordinates of space-time events in one inertial frame S,
  to observations of the coordinates
 ($x'$,$y'$,$z'$,$t'$) of the same events in another inertial frame S'.
   As is conventional, the Cartesian spatial coordinate
   axes of the two frames are parallel, and the origin of the frame S' moves with constant speed, $v$,
    along the $x$-axis.
   In any actual experiment, times are recorded by clocks, and positions specified by marks on fixed
    rulers (or their equivalent). Therefore, in order to relate the space-time coordinates appearing in the
   LT to actual physical measurements they must be identified with clock readings and length interval
   measurements~\cite{JHFSTP1}. This can be done in two distinct ways depending on whether the experimenter observing the
   clocks and performing the length measurements is at rest in S or in S'. In the former case (only events
   with spatial coordinates along the $x$-axis are considered so that $y$ = $y'$ = $z$ = $z'$ = $0$) the
   appropriate LT, for a clock, C', situated at the origin of S', is:
    \begin{eqnarray}
  x'(\rC')& = & \gamma_v[x(\rC')-v\tau] = 0 \\
 t'& = & \gamma_v[\tau-\frac{\beta_v x(\rC')}{c}]
\end{eqnarray}
 and in the latter case, for a clock, C, situated at the origin of S, is:
    \begin{eqnarray}
  x(\rC) & = & \gamma_v[x'(\rC)+c\tau'] = 0 \\
 t& = & \gamma_v[\tau'+\frac{\beta_v x'(\rC)}{c}] 
\end{eqnarray}
 In these equations $\beta_v \equiv v/c$, $\gamma_v  \equiv 1/ \sqrt{1-\beta_v^2}$ and $c$ is the speed of
 light in vacuum. The clocks in S and S' are synchronised so that for (2.1) and (2.2),
  $t'= \tau = 0$ when $x = x'= 0$, and for (2.3) and (2.4),
  $t = \tau' = 0$ when $x = x'= 0$.
    In (2.1) and (2.2) the transformed events lie on the worldline of a clock, C', at rest in S',
   which is observed from S. The observed time in S registered by C'( which is in motion in this frame)
  is $t'$, while $\tau$ is the time registered by the clock, C, identical to C', but at rest in S. In contrast,
  in (2.3) and (2.4) the transformed events lie on the worldline of C, which is observed from S'. The time
    $t$ is that registered by C as observed from S' and $\tau'$ is the time registered by C' as observed
   in its own proper frame. Thus two distinct experiments are possible involving one stationary and one moving
    clock, depending on whether the experimenter performing the space and time measurements is in the rest
   frame of one, or the other, of the two clocks. To describe both of these experiments, four different
    time symbols, $\tau$, $\tau'$, $t$ and $t'$, with different operational meanings, are required.
 \par From (2.1), the equation of motion of C' in S is:
   \begin{equation}
     x(\rC') = v \tau
   \end{equation}
  while from (2.3) the equation of motion of C in S' is:
  \begin{equation}
     x'(\rC) = -v \tau'
   \end{equation}
  Using (2.5) to eliminate $x$ from (2.2), and in view of the definition of $\gamma_v$:
       \begin{equation}
      \tau = \gamma_v  t'
  \end{equation}  
  Similarly, using (2.6) to eliminate $x'$ from (2.4) gives:
     \begin{equation}
     \tau' = \gamma_v  t
  \end{equation}  
    (2.7) and (2.8) are expressions of the relativistic Time Dilatation (TD) effect in the two
    `reciprocal' experiments that may be performed using the clocks C and C'. They show that, according,
    to the LT, `moving clocks run slow' in a universal manner (no spatial coordinates appear in (2.7) and (2.8)).
     In fact: 
   \begin{equation}
  \frac{{\rm rate~of~moving~clock}}{{\rm rate~of~stationary~clock}} =\frac{ t'}{\tau}
     = \frac{t}{\tau'} =\frac{1}{\gamma_v}
   \end{equation}
   \par To discuss measurements of the spatial separations of moving clocks, at least two clocks,
    (say, $\CAP$ and $\CBP$, at rest in S') must be considered. It is assumed that they lie along the $x'$-axis
     separated by the distance $L'$, $\CAP$ being at the origin of S' and $\CBP$ at $x' = L'$. The space transformation
    equations analogous to (2.1) for $\CAP$ and $\CBP$ and are then:
     \begin{eqnarray}
  x'(\CAP) & = & \gamma_v[x(\CAP)-v\tau] = 0 \\
  x'(\CBP)-L' & = & \gamma_v[x(\CBP)-L-v\tau] = 0   
\end{eqnarray}
   Inspection of (2.11) shows that $L = x(\CBP,\tau = 0)$, a relation valid for
   all values of $v$ for the choice of coordinate systems in (2.10) and (2.11).
     In particular, it is valid when $v \rightarrow 0$, $\gamma_v  \rightarrow 1$, and $x \rightarrow x'$. Then
       for $v = 0$:
   \begin{equation}
  x'(\CBP)-L' =  x'(\CBP)-L
    \end{equation}
     so that  
   \begin{equation}
  L'  =  L
    \end{equation}
  The spatial separation of the clocks is therefore a Lorentz-invariant quantity.
   \par Suppose now that two objects move with speeds $u_1$, $u_2$ ($u_1 > u_2$) along the positive 
   $x$-axis in S and that they are coincident with the origins of S and S' at the time $\tau = 0$. At later times
   $\tau$, $t'$, the separation of the objects in S is $(u_1-u_2)\tau$ and in S' is $(u'_1-u'_2) t'$.
   where  $u'_1 - u'_2$ is the relative velocity of the objects in S'. In view of (2.13)
   and the time dilatation formula (2.7) it follows that
      \begin{equation}  
  u'_1 - u'_2 = \gamma_v (u_1-u_2)
 \end{equation}
 A particular case of (2.14), to be used in the following section, is  $u_1 = v$, $u_2 = u'_1 = 0$ giving
  \begin{equation} 
  -u'_2 \equiv v' = \gamma_v v
  \end{equation}
  where $v'$ is the observed speed of the origin of S along the negative $x$-axis in S´.
   The relation (2.14) is the
   transformation formula of the {\it relative velocity} of two objects between two inertial frames, to be 
   contrasted with the relativistic parallel velocity addition formula:
  \begin{equation} 
   w = \frac{u-v}{1-\frac{uv}{c^2}}
  \end{equation}
   which relates kinematical configurations of a single moving object in the frames S and S'.
    In the case $u = 0$, (2.16) gives
     $w = -v$ and this equation relates the kinematical configuration in S in the primary experiment
    described by (2.1) and (2.2) to that in S' in the (physically independent) reciprocal  experiment
   described by (2.3) and (2.4).
     In contrast (2.14) describes the observed {\it relative velocity transformation} within the primary 
     experiment. For further discussion of this important point see Refs.~\cite{JHFSTP3,JHFRECP}.

 \SECTION{\bf{Muons are clocks that demonstrate time dilatation and differential aging}}
    Muon decays constitute an excellent laboratory for testing the predictions of special relativity.
    For example, the TD effect of Eqn(2.7) was experimentally
    confirmed at the per mille level of relative precision in the ultrarelativisic domain 
    ($\gamma_v \simeq 30$) by observation of the decay of muons in the
    storage ring of the last CERN muon $g-2$ experiment~\cite{NatureTD}.
    In the present paper, it is shown that thought experiments involving muons provide a graphic illustration
    of the predicted space-time behaviour, in special relativity, of clocks in different inertial frames.
    \par  Unlike most other unstable particles, muons are particularly suitable for
       precise tests of the TD effect because of the ease of their production from pion
       decay and their long mean lifetime of 2.2 $\mu$s. The former yields high events statistics and the latter
       the possibility of precise time interval measurements using accurate clocks in the
       laboratory frame~\cite{NatureTD}.
       \par The thought experiment developed in the present paper is an elaboration of the well-known
        demonstration that the very presence of cosmic muons at the Earth's surface is, by itself,
     sufficient to demonstrate the existence of the TD effect~\cite{FL,TW,TL,CC}. Muons are produced predominantly
      by the weak decay of charged pions $\pi^{\pm} \rightarrow \mu^{\pm} \nu$. The velocity of the muon, 
      $v_{\mu}$, depends upon that of the parent pion, $v_{\pi}$, and
    the centre-of-mass decay angle, $\theta^{\star}$.
      If the pion has the same velocity, $v_{\mu}^{\star} = c(m_{\pi}^2-m_{\mu}^2)/(m_{\pi}^2+m_{\mu}^2)$,
      as the muon in the pion rest frame, (corresponding to a pion momentum of 49.5 MeV/c)
      and $\cos \theta^{\star} = -1$, the muon is produced at rest in the laboratory system. 
      The maximum muon decay energy $E_{\mu}^{max}$ correponds to  $\cos \theta^{\star} = 1$ and is given, 
     in terms of the parent pion energy $E_{\pi}$, and the pion velocity $v_{\pi} = c \beta_{\pi}$, by the
      relation:
      \begin{equation}
     E_{\mu}^{max} = E_{\pi} \frac{[m_{\pi}^2(1+\beta_{\pi})+m_{\mu}^2(1-\beta_{\pi})]}{2 m_{\pi}^2}
       \end{equation} 
   
      For ultra-relativistic parent pions with $\beta_{\pi} \simeq 1$, $ E_{\mu}^{max} \simeq  E_{\pi}$.
     \par Due to the thickness of the Earth's atmosphere, the majority of interactions of primary
    cosmic protons, that produce the parent pions of cosmic muons, occur at high altitude, $\simeq$ 20 km above the Earth's surface. A muon with 
      speed close to that of light then takes at least $\simeq$ 700 $\mu s$ to reach the surface of the Earth.
     This may be compared with the muon mean lifetime of 2.2 $\mu s$. Without the TD effect, only
      a fraction $\exp[-700/2.2] \simeq 10^{-138}$ of the muons produced at altitude would reach
      the Earth's surface. However a 10 GeV muon, which has $\gamma_v \simeq 94$, has a 3.5 $\%$
       probability to
      reach the  Earth's surface, before decaying, when the TD effect is taken into account. 
      \par In the thought experiment considered here it is assumed that two muons \MA and \MAP
       are produced simultaneously at the same point A (see Fig.1a) by decay of pions from a primary cosmic
    ray interaction with the nucleus of a gas atom in the atmosphere. The muon \MA is produced at rest in
    the atmosphere (inertial frame S) while \MAP (with proper frame S') is produced with velocity $v = c \beta_v = \sqrt{3}/2$, so that
     $\gamma_v = 2$. It happens that both muons decay after time $T$ in their proper frames.
      Because of the TD effect, the muon \MAP will then be seen by an observer at rest in
     the atmosphere to decay after time $\tau = \gamma_v T = 2T$ at a point B at a distance 
     $L = 2Tv = 2.28$km from A. It is also supposed that at the same time, $\tau = 0$,  that \MA and \MAP
     are created, another muon, \MB, (also with proper decay lifetime $T$) is created at rest in the
      atmosphere at the 
    point B, by decay of pion from another primary cosmic ray interaction. Since \MA and \MB are at rest in 
     the atmosphere and have no TD effect, they will decay
    simultaneously at $\tau = T$ (Fig.1b) in the frame S. At this instant the muon \MAP is still undecayed and is
    at the point M, midway between A and B, When \MAP decays (Fig.1c)  \MA and \MB no longer
    exist, however the centres of mass of their, by now distant, decay products $e$, $\nu$ and $\bar{\nu}$,
     denoted as (\MA) and (\MB), and indicated in Figs. 1-3 as two concentric circles, still remain at the points A and B.

\begin{figure}[htbp]
\begin{center}\hspace*{-0.5cm}\mbox{
\epsfysize15.0cm\epsffile{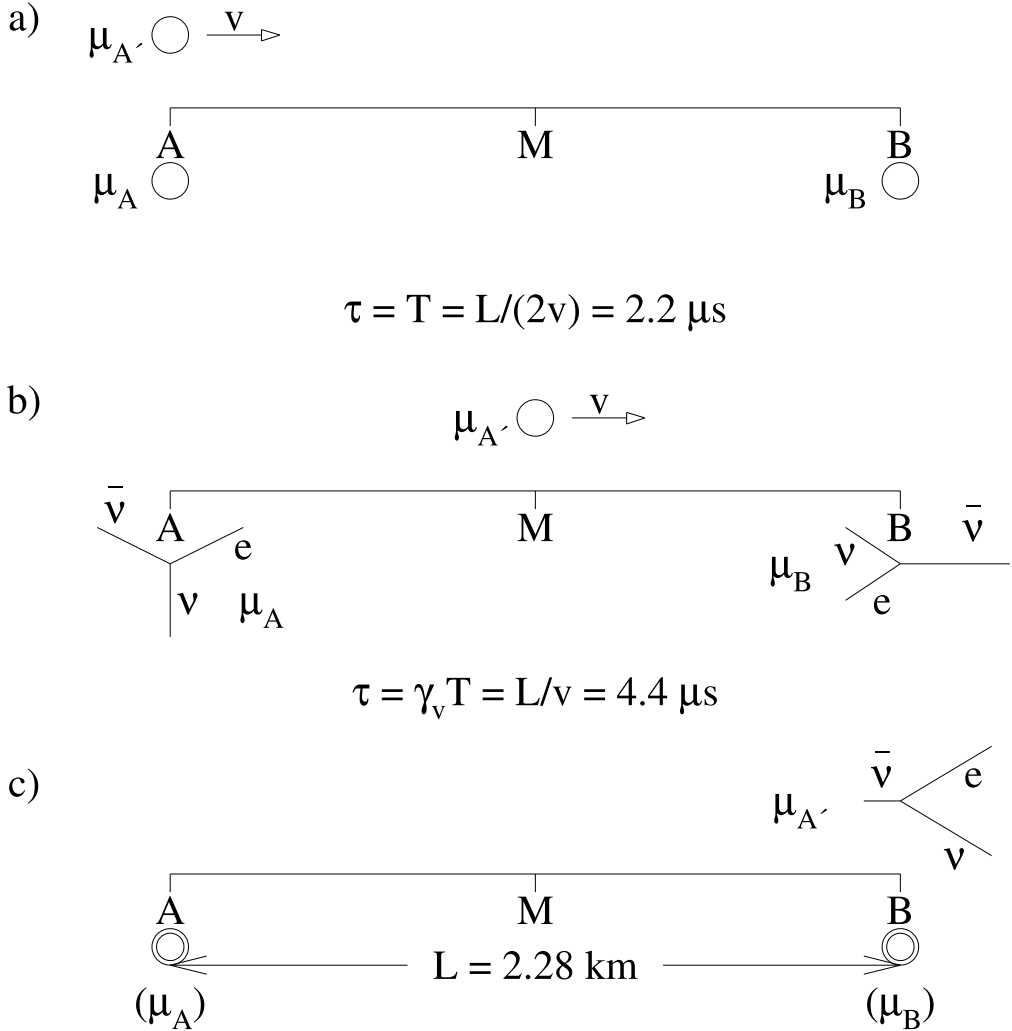}}
\caption{ {\em  The sequence of muon decay events as observed from the atmosphere (frame S).
  a) Muons \MAP, \MA and \MB are simultaneously created. Muon \MAP moves to the right with
 velocity $v = (\sqrt{3}/2)c$. b) At time $\tau = T$, muons \MA and \MB decay simultaneously.
  At this time \MAP is observed from S to be aligned with the mid-point, M, of A and B.
  c) At time $\tau = \gamma_v T$, muon \MAP is observed to decay. At this time it is at B,
 the centre of mass of the decay products of \MB. For clarity, the muons are shown displaced vertically.}}  
\label{fig-fig1}
\end{center}
\end{figure}

\begin{figure}[htbp]
\begin{center}\hspace*{-0.5cm}\mbox{
\epsfysize15.0cm\epsffile{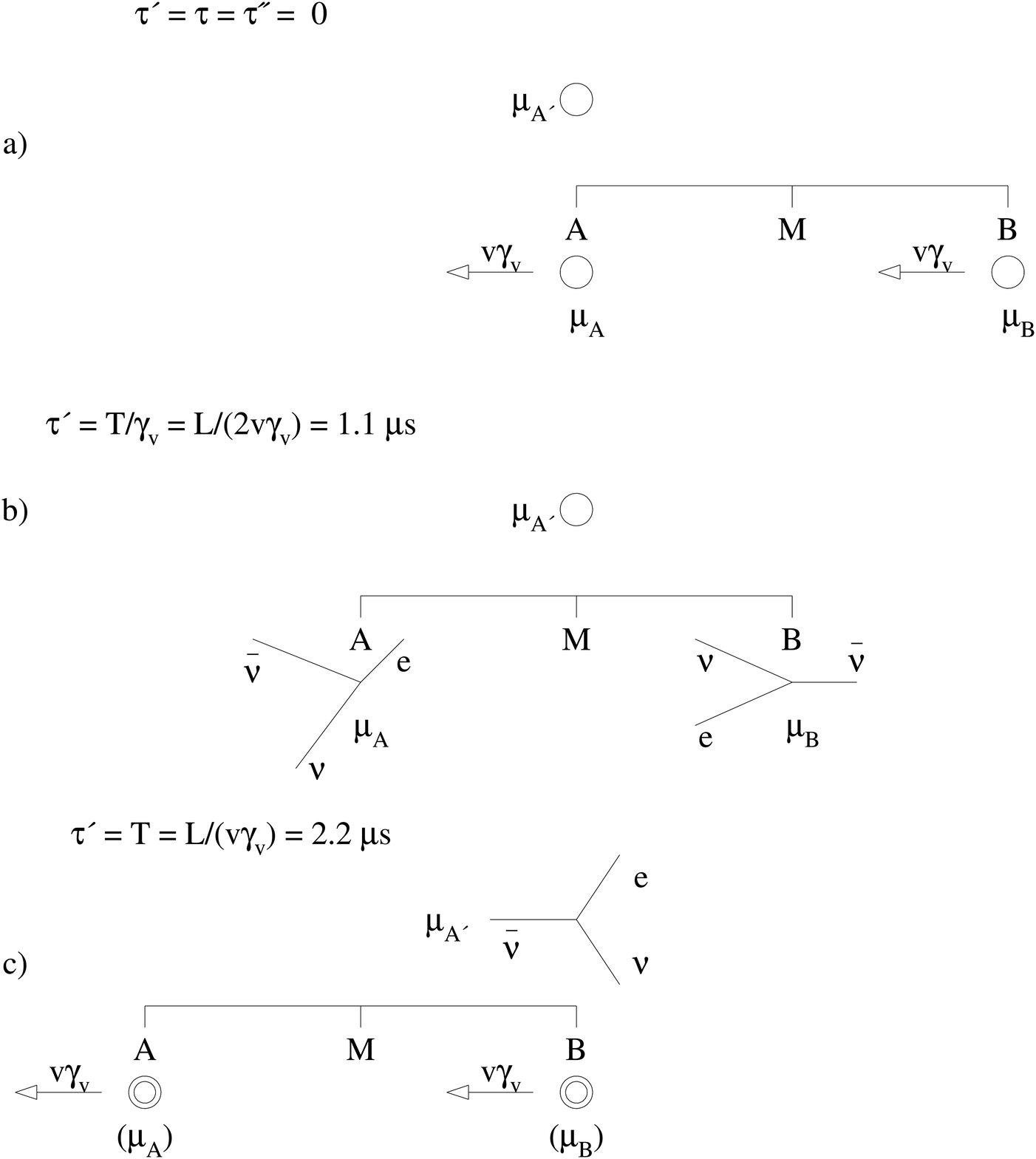}}
\caption{{\em The sequence of muon decay events as observed in the proper frame (S') of \newline \MAP.
  a) Muons \MAP, \MA and \MB are simultaneously created. Muons \MA and \MB are observed to
   move to the left with
 velocity $v \gamma_v = \sqrt{3}c$. b) At time $\tau' = T/\gamma_v$, \MA and \MB decay simultaneously.
  At this time, as in Fig.1, \MAP is aligned with the mid-point, M, of A and B.
  c) At time $\tau' =  T$ muon \MAP decays. At this time, as in Fig.1, it is aligned with B,
 the centre of mass of the decay products of \MB. For clarity, the muons are shown displaced vertically.}}  
\label{fig-fig2}      
\end{center}
\end{figure}

\begin{figure}[htbp]
\begin{center}\hspace*{-0.5cm}\mbox{
\epsfysize15.0cm\epsffile{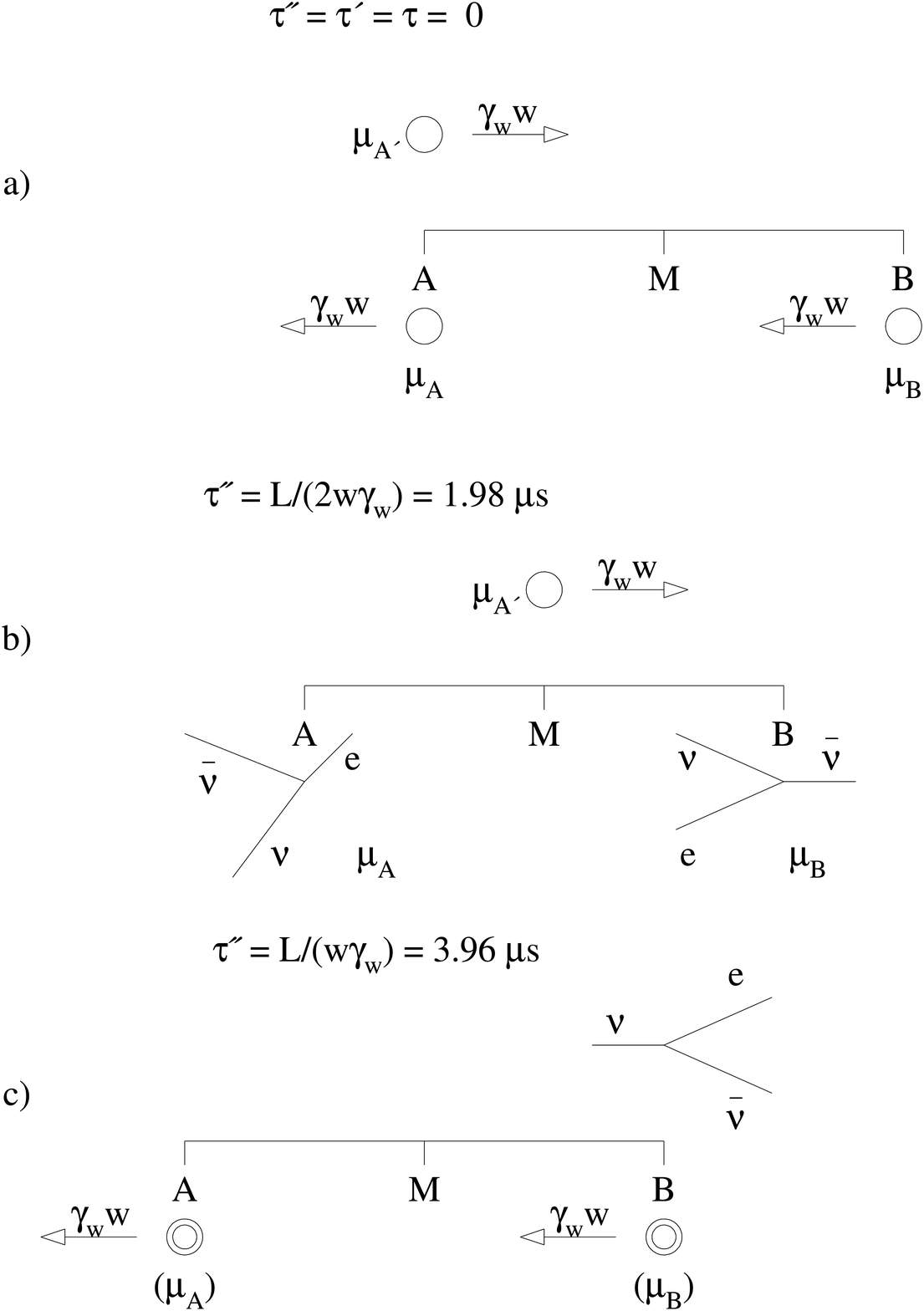}}
\caption{{\em  The sequence of muon decay events as observed from the frame (S'') that
  moves in S, parallel to the direction of motion of \MAP, with the velocity $w = v/2 = c\sqrt{3}/4$.
 a) Muons \MAP, \MA and \MB are simultaneously created. Muon \MAP is observed to move to the right with
  velocity $w \gamma_w$ and \MA and \MB  move to the left with
 velocity  $w \gamma_w$. b) At time $\tau'' = L/(2w\gamma_w)$, \MA and \MB decay simultaneously. At this time,
 as in Figs.1 and 2, \MAP is aligned with the mid-point, M, of A and B. c) At time  $\tau'' = L/(w\gamma_w)$
  \MAP decays. As in Figs.1 and 2 it is aligned, at with this time, with B, the centre of mass of the
   decay products of \MB. For clarity, the muons are shown displaced vertically.}}
\label{fig-fig3}
\end{center}
\end{figure}
    
 \par The sequence of events that would be seen by an observer in the rest frame, S,' of \MAP 
  corresponding to those of Fig.1, is shown in Fig.2. According to the relative velocity transformation
  formula (2.15),  \MA and \MB are observed to move to the left with velocity $v\gamma_v$. The
  configuration at $\tau = \tau' = 0$ is shown in Fig.2a. According to the time dilatation relation (2.7),
  appropriate to this case, \MA and \MB are seen to decay simultaneously at time
   $\tau' = T/\gamma_v = 1.2\mu s$ (Fig.2b). As in Fig.1, it can be sees that \MAP is aligned with M,
   the mid point of
   the line segment AB, at the time of simultaneous decay of \MA and \mbox{\MB.}  As shown in Fig.2c, 
    \MAP decays at time  $\tau' = T = 2.2\mu s$, when it is aligned with B, as is also the case for 
   an observer in the frame S, as shown in Fig.1c. Since \MA and \MB are seen to decay simultaneously 
    by observers at rest in both S and S', there is no RS  effect.
   \par The sequence of events that would be observed in the frame S'', moving with velocity $w$ in
   the positive $x$-direction will now be considered. According to the relative velocity transformation
    formula (2.14), \MAP is observed to move with speed $\gamma_w(v-w)$ in the positive $x$-direction
   in S'' since the {\it relative velocity} of S'' and S' in the frame S is $v-w$. Also, according
   to (2.15),  \MA and \MB are observed to move with speed  $w \gamma_w$ in the negative $x$-direction
   in S''. The velocity of \MAP relative to \MA and \MB in the frame S'' is then  $v \gamma_w$.
    The sequence of events seem by an observer at rest in S'' is illustrated, for the case $w = v/2$,
   in Fig.3. Using the time dilatation realation (2.7), with $v$ set equal to $w$ in order to relate
 times in S and S'', \MA and \MB decay at time $\tau'' = T/\gamma_w$, when \MAP is aligned with M
  (see Fig.3b). Note that the time of this event in S'' depends only on the value of $w$, being independent
  of the value of $v$. The muon \MAP decays at time $\tau'' = \gamma_v T/\gamma_w$ when it is aligned
   with B (see Fig.3c).
    \par   In all three frames, S, S' and S'',  \MA and \MB are observed to decay simultaneously 
    and earlier than \MAP. These decay times, in each frame, are presented in Table 1. The
    entries in this table satisfy the following condition:
   \begin{equation}
    \frac{\tau_D(\mu_{\rm A'})}{\tau_D(\mu_{\rm A})} = \frac{\tau'_D(\mu_{\rm A'})}{\tau'_D(\mu_{\rm A})}
     = \frac{\tau''_D(\mu_{\rm A'})}{\tau''_D(\mu_{\rm A})} = \gamma _v
    \end{equation}
     Since the last member of this equation is independent of $w$, it follows that  ratio of
    the decay times given by the time dilatation relation (2.7) is the same for all inertial
    observers, and so is an invariant, fixed by the velocity of \MAP in the rest frame of \MA 
    where the time dilatation effect is defined ---i.e. observer at rest in S, observed muon 
    at rest in S'.
    \par Different (but reciprocal, i.e. related to those of Eqn(3.2) by exchange of primed
    and unprimed quantities) results would be obtained in the situation where the observer of the
    time dilatation effect is at rest in S', while the observed muon is a rest in S, so that
    the reciprocal time dilatation relation (2.8) is applicable.
    \par Inspection of, and reflection on, Figs.1 and 2 reveals the physical basis of the differential
    aging effect in the `twin paradox' introduced by Langevin~\cite{Langevin}. The `travelling twin' can
    be indentified with \MAP, the `stay at home' one with either \MA or \MB since their associated
    `clocks' are synchronous. At the end of the outward journey, when  \MAP arrives at B, the 
    synchronous clocks at A and B were seen in S to run twice as fast as that of \MAP ---in fact 
     \MA and \MB have already decayed when \MAP arrives at B. In Fig.1, the observer in S sees
     \MAP aging less rapidly than  \MA or \MB. However, as required by the time dilatation relation
     (2.7), an observer in S' (Fig.2) sees  \MA or \MB {\it aging more rapidly}, by a factor two,
     than \MAP. This is true even though these clocks are in motion relative to the observer
      on S'. Fig.2 also reveals that the physical basis of time dilatation, and differential aging, is not, as hitherto,
      to be found in
    `length contraction' in the frame S', but instead in the greater velocity of \MA or \MB relative to \MAP 
      in the `travelling frame' S', than that of  \MAP  relative to  \MA or \MB in the `base frame'
      S from which the time dilatation effect is observed. For futher discussion of base and travelling
      frames in relation to primary and reciprocal space-time experiments see
      Refs.~\cite{JHFSTP3,JHFRECP}.
      \par The much-discussed `twin paradox' arises when it is attempted to describe the sequence of events
      shown in Fig.2 by use of the time dilatation relation (2.8) of the reciprocal experiment which requires
      clocks in S to run slower (not faster, as in Fig.2) than those in S', when observed in the latter
       frame. This is a nonsensical interpretation since the time variables $\tau$ and $t'$ appearing in the
      time dilatation relation (2.7) have a completely different operational meaning to those, 
       $\tau'$ and $t$ in the time dilatation relation of the reciprocal experiment. See Ref.~\cite{JHFSTP3}
       for a  more detailed discussion of the standard and incorrect interpretation of the twin paradox
      based on the spurious LC effect, in contrast with its correct interpretation following from
      the relative velocity transformation formula (2.14).

  \begin{table}
\begin{center}
\begin{tabular}{|c|c c c|} \hline
 Frame   &\multicolumn{1}{c|}{ $\tilde{\tau}_D(\mu_{{\rm A'}})$}
    &\multicolumn{2}{c|}{ $\tilde{\tau}_D(\mu_{{\rm A}}) = \tilde{\tau}_D(\mu_{{\rm B}})$}  \\ \hline
 S &\multicolumn{1}{c|}{ $\gamma_v T$}
    &\multicolumn{2}{c|}{$T$}  \\
S' &\multicolumn{1}{c|}{$T$}
    &\multicolumn{2}{c|}{$T/\gamma_v$}  \\
S'' &\multicolumn{1}{c|}{$\gamma_vT/\gamma_w$}
    &\multicolumn{2}{c|}{$T/\gamma_w$}  \\ 
 \hline 
\end{tabular}
\caption[]{{\em Decay times of muons \MAP and  \MA and \MB in the frames S, S' and S''.
  $\tilde{\tau}$ denotes the proper time in each frame.}}      
\end{center}
\end{table}

\pagebreak

\end{document}